\documentclass[twocolumn,showpacs,preprintnumbers,amsmath,amssymb,aps,prc,superscriptaddress]{revtex4-1}
\usepackage{graphicx}% Include figure files
\usepackage{dcolumn}% Align table columns on decimal point
\usepackage{bm}% bold math

\begin{document}

%\preprint{}

\title{Mass of astrophysically relevant $^{31}$Cl and \\the breakdown of the isobaric multiplet mass equation}


\author{A. Kankainen}
\email[]{anu.kankainen@jyu.fi}
\author{L. Canete}
\author{T. Eronen}
\author{J. Hakala}
\author{A. Jokinen}
\author{J. Koponen}
\author{I.D. Moore}
\author{D. Nesterenko}
\author{J. Reinikainen}
\author{S. Rinta-Antila}
\author{A. Voss}
\affiliation{University of Jyvaskyla, P.O. Box 35, FI-40014 University of Jyvaskyla, Finland}
\author{J. \"Ayst\"o}
\affiliation{Helsinki Institute of Physics, P.O. Box 64, FI-00014 University of Helsinki, Finland}
%\altaffiliation{University of Jyvaskyla, P.O. Box 35, FI-40014 University of Jyvaskyla, Finland}


\date{\today}

\begin{abstract}
The mass of $^{31}$Cl has been measured with the JYFLTRAP double Penning trap mass spectrometer at the Ion Guide Isotope Separator On-Line (IGISOL) facility. The determined mass-excess value, $-7034.7(34)$~keV, is 15 times more precise than in the Atomic Mass Evaluation 2012. The quadratic form of the isobaric multiplet mass equation for the $T=3/2$ quartet at $A=31$ fails ($\chi^2_n=11.6$) and a non-zero cubic term, $d=-3.5(11)$~keV, is obtained when the new mass value is adopted. $^{31}$Cl has been found to be less proton-bound with a proton separation energy of $S_p=265(4)$~keV. Energies for the excited states in $^{31}$Cl and the photodisintegration rate on $^{31}$Cl have been determined with significantly improved precision using the new $S_p$ value. The improved photodisintegration rate helps to constrain astrophysical conditions where $^{30}$S can act as a waiting point in the rapid proton capture process in type I x-ray bursts.
\end{abstract}

%\keywords{}

\maketitle

$^{31}$Cl is a short-lived ($T_{1/2}=190(1)$~ms \cite{Saa11}) $sd$-shell nucleus and a well-known beta-delayed proton emitter \cite{Ays82,Ays83,Ogn96,Kan06,Saa11}. However, its mass-excess value ($\Delta=-7066(50)$~keV \cite{AME12}) is still based on a single $Q$-value measurement of $^{36}$Ar$(^3$He$,^8$Li$)^{31}$Cl reaction performed at Michigan State University in the 1970s \cite{Ben77}. The mass of $^{31}$Cl is relevant for testing the isobaric multiplet mass equation (IMME) \cite{Wig57,Wei59} as it is a member of the $T=3/2$ isobaric quartet with isospin projection $T_Z=(N-Z)/2=-3/2$. According to the IMME, the masses of the isobaric analog states (IAS) in a mass multiplet should show purely quadratic behavior: $M(A,T,T_Z)=a(A,T)+b(A,T)T_Z+c(A,T)T_Z^2$ after treating the Coulomb interaction using the first-order perturbation theory. Previous IMME evaluations have shown that the quadratic form works well for the $T=3/2$ quartet at $A=31$ \cite{Ben79,Bri98,Lam13,Mac14,Mac14err} but the test has not been very stringent, mainly due to the uncertainty in the $^{31}$Cl mass. Overall, the quadratic form of the IMME has failed only in a few cases, such as at $A=8$ \cite{Cha11}, $A=9$ \cite{Bro12}, $A=21$ \cite{Gal14}, $A=32$ \cite{Kan10,Kwi09}, $A=35$ \cite{Yaz07}, and $A=53$ \cite{Zha12}. The breakdown of the IMME has been explained, e.g. by isospin mixing of the states and charge-dependent effects \cite{Bro12,Sig11}. However, for some cases, such as for the $A=53$ quartet \cite{Zha12}, even detailed shell-model calculations have not been able to describe the breakdown. 

The mass of $^{31}$Cl is also relevant for the rapid proton capture ($rp$) process occurring in type I x-ray bursts (XRB) \cite{Fis04,Fis08}. There, most of the nucleosynthetic flow proceeds through $^{30}$S which can act as a waiting point due to its half-life ($1.178(5)$~s \cite{Wil80}) and low proton-capture $Q$ value establishing a $(p,\gamma)-(\gamma,p)$ equilibrium towards $^{30}$S at high temperatures. The route via the $^{30}$S$(\alpha,p)^{33}$Cl reaction is hindered by the Coulomb barrier at typical XRB temperatures of around 1 GK. Waiting points, such as $^{30}$S, have been proposed to be responsible for the double-peaked structure observed in XRB luminosity curves \cite{Fis04}.

The proton captures on $^{30}$S are dominated by resonant captures to the two lowest excited states in $^{31}$Cl. These have been studied via beta-delayed proton decay of $^{31}$Ar \cite{Axe98,Axe98err,Fyn00,Kol14} with observed laboratory energies of 446(15) and 1415(5) keV \cite{Axe98} and 1416(2) keV \cite{Fyn00}. Recently, $^{31}$Cl has been studied via Coulomb breakup of $^{31}$Cl at high energy in inverse kinematics using the R$^3$B-LAND setup at GSI \cite{Lan14}. The two lowest-lying levels, $1/2^+$ at 782(32)~keV and $5/2^+$ at 1793(26) keV \cite{Lan14}, were found to be in a reasonable agreement with the estimations, 745(17) and 1746(9) keV \cite{Wre09}, based on the IMME and beta-delayed proton data. However, also lower excitation energies, 726(37) keV and 1731(82) keV, have been reported from R$^3$B-LAND \cite{LangerPhD}. A similar Coulomb dissociation study of $^{31}$Cl performed at RIKEN resulted in resonance energies of around 0.45 and 1.3 MeV \cite{Tog11}. In order to compare the results from R$^3$B-LAND with the beta-delayed proton data, and to verify the excitation energies of the lowest resonance states in $^{31}$Cl, the proton separation energy of $^{31}$Cl, \emph{i.e.} its mass, has to be known more precisely. 

To estimate the waiting-point conditions for $^{30}$S, also the rate for photodisintegration reactions on $^{31}$Cl ($\lambda_{\gamma,p}$) has to be taken into account. The ratio of $\lambda_{\gamma,p}$ to the proton-capture reaction rate $N_A\left\langle \sigma v\right\rangle$ depends exponentially on the proton-capture $Q$ value on $^{30}$S (\emph{i.e.} the proton separation energy $S_p$ of $^{31}$Cl) \cite{Ili07}:

\begin{equation}
\label{eq:gprate}
\begin{split}
\frac{\lambda_{\gamma,p}}{N_A\left\langle \sigma v\right\rangle}&=9.8685\times10^9T_9^{3/2}\frac{g_{S}g_p}{g_{Cl}}\left(\frac{G_SG_p}{G_{Cl}}\right)\\
&\times \left(\frac{m_Sm_p}{m_{Cl}}\right)^{3/2}e^{-11.605Q/T_9},
\end{split}
\end{equation}
where $m_i$ are the masses in atomic mass units, $g_i$ the statistical factors $g_i=2J_i+1$ and $G_i$ normalized partition functions for $^{30}$S, $p$ and $^{31}$Cl. The normalized partition functions \cite{Rau00} are close to one in the relevant energy region. The uncertainty in the present $Q$ value has been shown to significantly affect XRB nucleosynthesis calculations in a high-temperature ($T_{peak}=2.50$~GK) scenario with normal burst duration ($\approx$100~s) as well as in a short burst ($\approx$ 10 s) scenario with $T_{peak}=1.36$~GK \cite{Par09}.

\begin{figure}
\centering
\includegraphics[width=0.45\textwidth,clip]{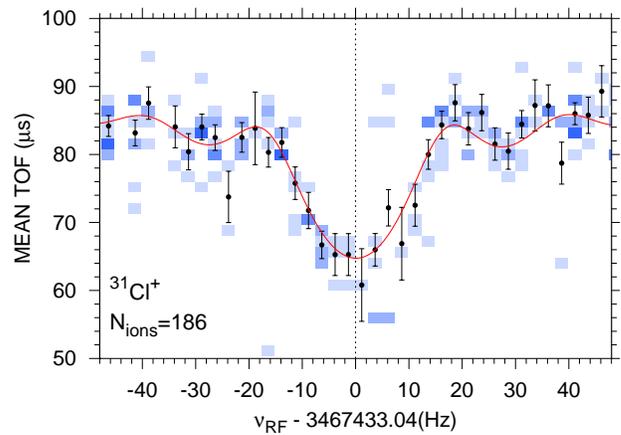}
\caption{(Color online) TOF-ICR spectrum of $^{31}$Cl$^+$ with a quadrupolar RF excitation of 50~ms. The spectrum represents a typical resonance of $^{31}$Cl obtained in 140~minutes. The blue squares indicate the number of ions in each time-of-flight bin: the darker the color, the greater the number of ions.}
\label{fig:tof}       
\end{figure}

\begin{figure}
\centering
\includegraphics[width=0.45\textwidth,clip]{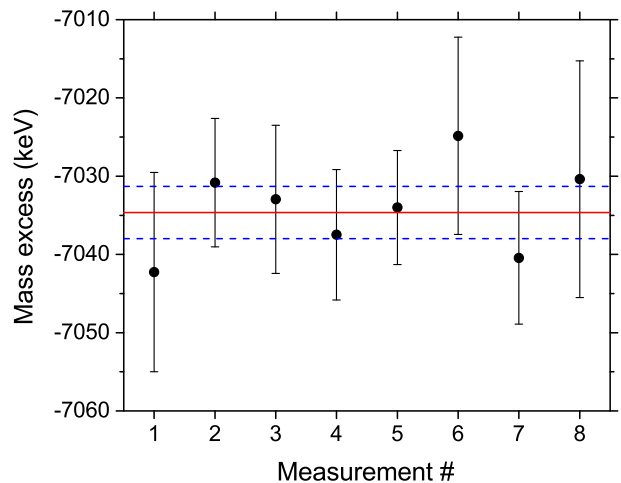}
\caption{(Color online) Mass-excess values determined in this work. The red line shows the weighted mean of the results and the dashed blue lines $1\sigma$ error bands.}
\label{fig:ME}       
\end{figure}

$^{31}$Cl$^+$ ions were produced via $^{32}$S$(p,2n)^{31}$Cl reactions using a 40-MeV proton beam impinging on a 1.8-mg/cm$^2$-thick ZnS target at the IGISOL facility \cite{Moo13}. The reaction products were stopped in helium gas and extracted with a sextupole ion guide \cite{Kar08} and accelerated to 30 keV before mass-separation with a 55$^\circ$ dipole magnet. A radiofrequency quadrupole cooler and buncher \cite{Nie01} was implemented to convert the continuous $A=31$ beam into short ion bunches which are released into the JYFLTRAP double Penning trap mass spectrometer \cite{Ero12}. Simultaneous magnetron and cyclotron excitations were applied for the ions in the purification trap for 40~ms to select the $^{31}$Cl$^+$ ions using the mass-selective buffer gas cooling method \cite{Sav91}. In the precision trap, a 10-ms magnetron excitation was followed by a short, 50-ms cyclotron excitation to minimize the decay losses of $^{31}$Cl. The ion's cyclotron resonance frequency $\nu_c=qB/(2\pi m)$, where $q$ and $m$ are the charge and mass of the ion, respectively, was determined using the time-of-flight ion cyclotron resonance (TOF-ICR) technique \cite{Kon95} (see Fig.~\ref{fig:tof}). The magnetic field strength $B$ was calibrated using $^{31}$P$^+$ ions as a reference ($m(^{31}$P$)=$30.9737619984(7)~u \cite{AME12}). Thus, the atomic mass of $^{31}$Cl was determined using $m(^{31}$Cl$)=r(m_{ref}-m_e)+m_e$, where $r=\frac{\nu_{c,ref}}{\nu_c}$ is the cyclotron frequency ratio of $^{31}$P$^+$ and $^{31}$Cl$^+$, $m_{ref}$ and $m_e$ are the $^{31}$P and electron masses, respectively. The weighted mean of the measured frequency ratios was $r=1.000 603 30(12)$ resulting in a mass-excess value $\Delta=-7034.7(34)$~keV (see Fig.~\ref{fig:ME}), which is 31 keV higher than the value in the Atomic Mass Evaluation 2012 (AME12) \cite{AME12}. The uncertainty is dominated by the statistical error of the frequency fit. The systematic uncertainties, as described in Ref.~\cite{Elo09}, have a negligible contribution to the final result. 

The IMME was studied at $A=31$ using the new mass value for $^{31}$Cl. The ground-state masses for the other members of the multiplet have been taken from AME12 \cite{AME12} (see Table~\ref{tab:IAS}). The mass values of $^{31}$S and $^{31}$P are based on Penning-trap measurements at JYFLTRAP \cite{Kan10} and the Florida State University trap \cite{Red08}. The mass of $^{31}$Si is linked via $(n,\gamma)$ measurements (see, e.g. Refs.~\cite{Isl90,Ram92,Rot97,Pau01}) to $^{29}$Si, which has been precisely measured with a Penning-trap at Massachusetts Institute of Technology \cite{Rai05}. The excitation energy for the $T=3/2$ IAS in $^{31}$S is based on data from beta-delayed $\gamma$-rays of $^{31}$Cl \cite{Kan06,Saa11} as well as from $^{31}$P$(^3$He$,t)$ \cite{Wre09a} and $^{33}$S$(p,t)$ reactions \cite{Nan79}. The energy for the IAS in $^{31}$P has been determined with high precision using the Gammasphere detector array \cite{Jen06}. A similar excitation energy, $E_x=6380.0(20)$~keV, has also been obtained via $^{30}$Si$(p,\gamma)$ measurements \cite{Wil67,Wol68,deNe75}. Thus, the data for the IMME are based on various independent measurements which do not show any significant discrepancies.

\begin{figure}
\centering
\includegraphics[width=0.45\textwidth,clip]{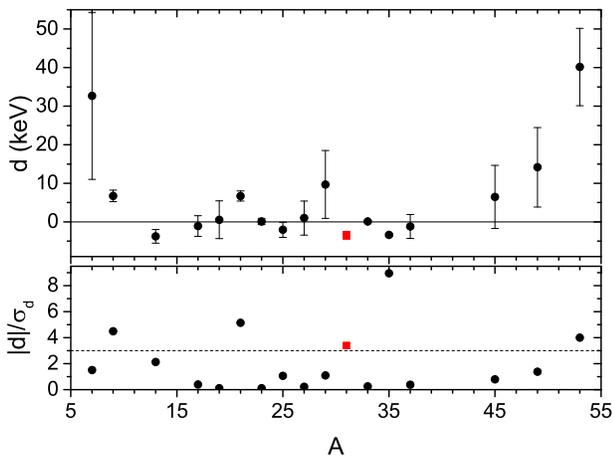}
\caption{Cubic coefficients for the known (lowest) $T=3/2$ isobaric quartets. The value for $A=31$ (red square) is from this work, for $A=21$ from Ref.~\cite{Gal14}, and the rest have been adopted from Ref.~\cite{Mac14}. The lower panel shows the significance of the deviation from zero.}
\label{fig:dcoeff}       
\end{figure}

Table~\ref{tab:coeff} summarizes the IMME fit results. With the new $^{31}$Cl mass value, the quadratic IMME fails ($\chi^2_n=11.6$) and a significant non-zero cubic coefficient $d=-3.5(11)$~keV is obtained. The more precise mass for $^{31}$Cl reveals the breakdown of the IMME: with the AME12 mass value for $^{31}$Cl the quadratic IMME fits perfectly well ($\chi^2_n=0.08$). So far, only $A=9$ \cite{Bro12}, $A=35$ \cite{Yaz07}, $A=53$ \cite{Zha12}, and recently $A=21$ \cite{Gal14}, of the known $T=3/2$ quartets have shown significant non-zero cubic coefficients (see Fig.~\ref{fig:dcoeff}). New precision measurements pave the way towards more fundamental understanding of the reasons behind the breakdown. Isospin mixing has successfully explained the breakdown of the IMME at $A=9$ \cite{Bro12} and $A=32$ \cite{Sig11} but failed at $A=21$ \cite{Gal14} albeit detailed shell-model calculations were carried out. 

The role of isospin mixing in the IMME is not straightforward. The quadratic IMME works well for the $A=33$, $J^\pi=1/2^+,T=3/2$ quartet ($\chi^2_n=0.06$ \cite{Mac14}) although isospin-forbidden beta-delayed protons observed from the IAS at 5548 keV in $^{33}$Cl (see, \emph{e.g.} Refs.~\cite{Hon96,Adi10}) imply there must be isospin mixing in the IAS. Interestingly, the cubic coefficients for the $A=31$ ($d=-3.5(11)$ keV) and $A=35$ ($d=-3.37(38)$ keV) $J^\pi=3/2^+,T=3/2$ quartets are very similar, which motivates further theoretical studies of these neighboring members of the $A=4n+3$ series of the $T=3/2$ quartets. Isospin mixing has been discussed for $A=35$ \cite{Yaz07,Ekm04} but no clear explanation for the breakdown has been given so far. Isospin mixing is plausible also for the $A=31$ quartet as there are candidates for the $T=1/2,3/2^+$ states \cite{Doh14} close to the $T=3/2$ IAS.

\begin {table}
\caption{\label{tab:IAS} Mass-excess values $\Delta$ and excitation energies $E_x$ for the $J^\pi=3/2^+,T=3/2$ isobaric analog states at $A=31$. The mass-excess value of $^{31}$Cl is from this work, the others are from the AME12 \cite{AME12}.}
\begin{ruledtabular}
\begin{tabular}{llll}
Nucleus  & $T_Z$ & $\Delta$ (keV) & $E_x$ (keV) \\
\hline
$^{31}$Cl & $-3/2$ & $-7034.7(34)$ & 0 \\
$^{31}$S & $-1/2$ & $-19042.52(23)$ & $6280.60(16)$ \cite{NDS31}\\
$^{31}$P & $+1/2$ & $-24440.5411(7)$ & $6380.8(17)$ \cite{Jen06}\\
$^{31}$Si & $+3/2$ & $-22949.04(4)$ & 0\\
\end{tabular}
\end{ruledtabular}
\end{table}

\begin {table}
\caption{\label{tab:coeff} Coefficients for the quadratic and cubic IMME fits (in keV) for the $T=3/2$ quartet at $A=31$. }
\begin{ruledtabular}
\begin{tabular}{lll}
  & Quadratic & Cubic \\
\hline
a & -15465.4(26)\footnotemark[1]  & -15463.2(10) \\
b & -5302.7(32)\footnotemark[1] 	& -5296.9(20) \\
c & 209.1(32)\footnotemark[1] 		& 209.5(10)\\
d & -  						& -3.5(11) \\
$\chi^2_n$ & 11.6 & - \\
\end{tabular}
\footnotetext[1]{The parameter uncertainty has been scaled with $\sqrt{\chi^2_n}$. }
\end{ruledtabular}
\end{table}

The breakdown of the IMME at $A=31$ is a crucial finding since the IMME prediction from Ref.~\cite{Saa11} has been used to establish level energies in $^{31}$Cl \cite{NDS31} from the beta-delayed proton data of $^{31}$Ar \cite{Axe98,Axe98err,Fyn00,Kol14}. The new mass value of $^{31}$Cl shows it is less bound than previously expected. The proton separation energy $S_p=265(4)$~keV is 31 keV lower and $\approx$13 times more precise than the AME12 value ($S_p=296(50)$~keV \cite{AME12}). The new mass measurement shifts all levels based on beta-delayed proton data \cite{Axe98,Axe98err,Fyn00,Kol14} 18 keV lower in energy and reduces the inherent systematic uncertainties from 50 keV to 4 keV. The revised energy for the $J^\pi=5/2^+,T=5/2$ IAS in $^{31}$Cl, the member of the $T=5/2$ sextet at $A=31$, is 12292.2(23)~keV based on Refs.~\cite{Axe98,Fyn00} and the $S_p$ and $S_{2p}$ values from this work. 

The two lowest excited states in $^{31}$Cl are relevant for the radiative resonant proton captures in the $rp$ process. By combining the new $S_p$ value with the beta-delayed proton data of Refs.~\cite{Axe98,Fyn00}, excitation energies of 726(16) and 1728(4) keV are obtained for the $1/2^+$ and $5/2^+$ states, respectively. These are about 15 keV lower than the presently recommended values (740(50) and 1746(5) keV \cite{NDS31}) and in a perfect agreement with the R$^3$B results 726(37) and 1731(82) keV \cite{LangerPhD}. The first excited state also agrees well with the USDB shell-model value of 724 keV \cite{Lan14}. The weighted mean for the resonance energies was calculated from Refs.~\cite{Axe98,Fyn00,LangerPhD} using the $S_p$ value from this work for Ref.~\cite{LangerPhD}. The values from Ref.~\cite{Lan14} deviate by $\approx 2\sigma$ and were not included. The resulting resonance energies, $E_r=461(15)$~keV and 1463(2)~keV, are very close to the beta-delayed proton data \cite{Axe98,Fyn00} and do not change the calculated proton-capture rates from Ref.~\cite{Wre09}.

\begin{figure}
\centering
\includegraphics[width=0.45\textwidth,clip]{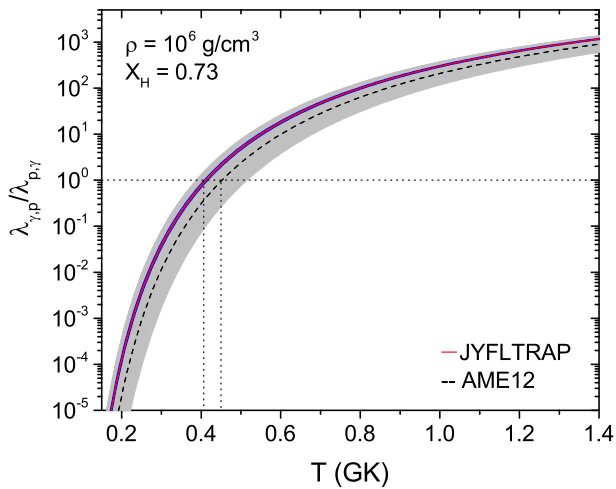}
\caption{(Color online) The ratio of $(\gamma,p)$ to $(p,\gamma)$ rates for typical XRB conditions. The uncertainties related to the JYFLTRAP $Q$ value are shown by the blue lines and to the AME12 value by the grey-shaded area.}
\label{fig:ratio}       
\end{figure}

The new $S_p$ value was used to compute the ratio of photodistegration rate on $^{31}$Cl to the proton capture rate on $^{30}$S according to Eq.~\ref{eq:gprate} and using $\lambda_{p,\gamma}=N_A\left\langle \sigma v\right\rangle_r\rho\frac{X_i}{m_{H}}$ for typical XRB conditions with density $\rho=10^6$ g/cm$^3$ and hydrogen mass fraction of $X_H=0.73$. The uncertainty related to the $Q$ value has been significantly reduced and photodisintegration rate takes over at lower temperatures compared to the ratio calculated with the AME12 $Q$ value (see Fig.~\ref{fig:ratio}). Above 0.44(1) GK, at least 20~\% of the reaction and decay flow has to wait for $\beta^+$ decay of $^{30}$S and it becomes a waiting point. The upper temperature limit for $^{30}$S waiting point, 1.0(3) GK, comes from the rate of the unmeasured $^{30}$S$(\alpha,p)^{33}$Cl reaction \cite{Wre09}.

The JYFLTRAP Penning-trap mass measurement of $^{31}$Cl has shown that the quadratic IMME fails at $A=31$ and the cubic term is non-zero. Theoretical calculations are anticipated to explain the deviation from the quadratic form and to explore possible underlying reasons for similarities in the cubic coefficients for $A=31$ and $A=35$. Isospin mixing between $T=1/2$ and $T=3/2$ states is plausible as there are candidates for $3/2^+$ states lying nearby the IAS. The improved precision in the proton separation energy of $^{31}$Cl has reduced the uncertainties related to excitation energies in $^{31}$Cl and the photodisintegration rate of $^{31}$Cl. Photodisintegration starts to dominate at lower temperatures than previously thought. The improved rate will be useful for future XRB model calculations helping to interpret the observed double-peaked structure in the luminosity curves.

\begin{acknowledgments}
This work has been supported by the Academy of Finland under the Finnish Centre of Excellence Programme 2012-2017 (Nuclear and Accelerator Based Physics Research at JYFL). A.K., D.N. and L.C. acknowledge the support from the Academy of Finland under project No. 275389.
\end{acknowledgments}

%\bibliography{cl31_bib}


\end{document}